\documentclass[12pt,a4paper,reqno]{amsart} 
\usepackage[T1]{fontenc} 
\usepackage{amsmath,amsxtra,amsfonts,amssymb,amsthm} 
\usepackage{a4}
\usepackage[vcentermath]{youngtab}
\usepackage{mathptm}

\DeclareSymbolFontAlphabet{\mathbm}{AMSb}

\newcommand{\sgn}{\operatorname{sgn}}

\newcommand{\menge}[1]{\mathbm{#1}} 
\newcommand{\eins}{\menge{I}} 
\newcommand{\SLC}{\ensuremath{\mathit{SL}(2,\menge{C})}}

\newcommand{\dif}{\ensuremath{\mathrm{d}}}

\newcommand{\Lcal}{\ensuremath{\mathcal{L}}} 
\newcommand{\D}{\ensuremath{\partial}}
\newcommand{\xbar}{\ensuremath{\widetilde{x}}} 
 
\newcommand{\Rd}{\ensuremath{\menge{R}^{d}}} 
\newcommand{\Rdon}{\Rd\setminus\{0\}} 
\newcommand{\Rv}{\ensuremath{\menge{R}^{4}}} 
 
\newcommand{\Rvm}{\ensuremath{\menge{R}^{4m}}} 
\newcommand{\Rvmon}{\ensuremath{\menge{R}^{4m}\setminus\{0\}}}

\newcommand{\Dcal}{\ensuremath{\mathcal{D}}}

\newcommand{\Rcal}{\ensuremath{\mathcal{R}}}
\newcommand{\Ccal}{\ensuremath{\mathcal{C}}} 
\newcommand{\Scal}{\ensuremath{\mathcal{S}}}

\newcommand{\Acal}{\ensuremath{\mathcal{A}}} 
 
\newcommand{\scp}[2]{\ensuremath{\left\langle #1, #2 \right\rangle}} 
 
\newcommand{\W}[2]{\ensuremath{W_{(#1;#2)}}} 
\newcommand{\tR}[2]{\ensuremath{t_{(#1;#2)}}}
\newcommand{\tiltR}[2]{\ensuremath{\overline{t}_{(#1;#2)}}} 
\newcommand{\tRli}[2]{\ensuremath{t_{(#1;#2)}^{\mathrm{linv}}}}

 \newcommand{\Xdot}{\dot{X}} 
\newcommand{\Ydot}{\dot{Y}}  
\newcommand{\Adot}{\dot{A}} \newcommand{\Bdot}{\dot{B}}

\newtheorem*{definition}{Definition}

\begin{document}
\title[Lorentz Covariance in Epstein-Glaser Renormalization]
{\Large Lorentz Covariance in Epstein-Glaser Renormalization}
\author[Dirk Prange]%
{\large Dirk Prange \\
\scriptsize\mbox{}\\ \mbox{} \\
II. Institut f\"ur Theoretische Physik\\
Universit\"at Hamburg\\
Luruper Chaussee 149\\
D-22761 Hamburg
Germany\\
email: \texttt{dirk.prange@desy.de}
}

\thanks{Work supported by the DFG Graduiertenkolleg ``Theoretische
  Elementarteilchenphysik'' Hamburg}

\bibliographystyle{amsalpha}

\begin{abstract}
  We give explicit and inductive formulas for the construction of a
  Lorentz covariant renormalization in the EG approach. This
  automatically provides for a covariant BPHZ subtraction at totally
  spacelike momentum useful for massless theories.
\end{abstract}

\maketitle

\section*{Introduction}
Our introduction is rather brief since this paper mainly is a
continuation of \cite{prep:wir}. There  we a gave
a formula for a Lorentz invariant renormalization in one
coordinate. We use the same methods to construct a covariant solution
for more variables. 

We review the EG subtraction and give some useful formulas in
section~\ref{sec:sub}. The cohomological analysis that leads to a group
covariant solution is reviewed in section~\ref{sec:g}. In
sections~\ref{sec:tensor},~\ref{sec:spinor} we give an inductive
construction for these solutions in the case of tensorial and
spinorial Lorentz covariance. Since the symmmetry of the one variable
problem is absent in general some permutation group calculus will
enter. Necessary material is in the appendix.  In the last
section~\ref{sec:bphz} we give a covariant BPHZ subtraction for
arbitrary (totally spacelike) momentum by fourier transformation.

\section{The EG subtraction}
\label{sec:sub}
We review the extension procedure in the EG approach \cite{pap:ep-gl}.
A more general introduction can be found in \cite{pap:prange1}, a
generalization to manifolds in \cite{proc:fred-brun,prep:fred-brun2}.
Let $ \Dcal^{\omega}(\Rd) $ be the subspace of test functions
vanishing to order $ \omega $ at $ 0 $. As usual, any operation on
distributions is defined by the corresponding action on testfunctions.
Define
\begin{gather}
\W{\omega}{w}:\Dcal(\Rd)\rightarrow\Dcal^{\omega}(\Rd), 
\quad\varphi\rightarrow\W{\omega}{w}\varphi, \notag \\
\left(\W{\omega}{w}\varphi\right)(x)=\varphi(x)-w(x)\sum_{|\alpha|\leq\omega} 
\frac{x^{\alpha}}{\alpha!}\D_{\alpha}\left(\varphi w^{-1}\right)(0),
\label{def:W}
\end{gather}
with $ w\in\Dcal(\Rd), w(0)\not=0 $. If ${^0t}$ is a distribution on
$\Dcal(\Rdon)$ with singular order $ \omega $ then ${^0t}$ can be
defined uniquely on $ \Dcal^{\omega}(\Rd) $, and
\begin{equation}
\scp{\tR{\omega}{w}}{\varphi}\doteq\scp{{^0t}}{\W{\omega}{w}\varphi}
\label{def:tR}
\end{equation}
defines an extension -- called renormalization -- of $^0t $ to the
whole testfunction space, $ \tR{\omega}{w}\in\Dcal'(\Rd) $. With the Leibnitz rule
\begin{equation}
\D_{\gamma}(fg)
=\gamma!\sum_{\mu+\nu=\gamma}\frac{1}{\mu!\nu!}\D_{\mu}f\,\D_{\nu}g,
\label{eq:leibnitz}
\end{equation}
we find
\begin{equation}
\D_{\gamma}\biggl(w\sum_{|\alpha|\leq\omega} 
\frac{x^{\alpha}}{\alpha!}\D_{\alpha}(\varphi w^{-1})(0)\biggr)(0) 
=\D_{\gamma}\varphi(0),
\label{eq:diffrest}
\end{equation}
for $ |\gamma|\leq\omega $, verifying the projector properties of
$W$. The $W$-operation (\ref{def:W}) is simplified if we require 
$w(0)=1$ and $\partial_{\alpha}w(0)=0$, for $0<|\alpha|\leq\omega$ 
(this was our assumption in \cite{prep:wir}). It can be achieved by 
the following $V$-operation ($\D_\mu w^{-1}$ means 
$\D_{\mu}(w^{-1})$):
\begin{equation}
V_{\omega}:\Dcal(\Rd)\mapsto\Dcal(\Rd),\quad (V_{\omega}w)(x)\doteq 
w(x)\sum_{|\mu|\leq \omega}\frac{x^{\mu}}{\mu!}\D_{\mu}w^{-1}(0),
\label{def:V}
\end{equation}
where $w(0)\not=0$ is still assumed.   We can write $W$ as
\begin{equation}
  \label{eq:varW}
\left(\W{\omega}{w}\varphi\right)(x)
=\varphi(x)-\sum_{|\alpha|\leq\omega} 
\frac{x^{\alpha}}{\alpha!}V_{\omega-|\alpha|}w\,\D_{\alpha}\varphi(0).
\end{equation}

The extension \eqref{def:tR} is not unique. We can add any polynomial 
in derivatives of $\delta$ up to order $\omega$:
\begin{align}
\scp{\tR{\omega}{w}}{\varphi}&=\scp{\tiltR{\omega}{w}}{\varphi}
+\sum_{|\alpha|\leq\omega} 
\frac{a^{\alpha}}{\alpha!}\D_{\alpha}\varphi(0), \\
\intertext{or rearranging the coefficients} 
&=\scp{\tiltR{\omega}{w}}{\varphi}
+\sum_{|\alpha|\leq\omega} 
\frac{c^{\alpha}}{\alpha!}\D_{\alpha}\left(\varphi w^{-1}\right)(0)
\label{def:tiltR}
\end{align}
Since $\W{\omega}{w}( w x^\alpha) = \W{\omega}{w} (x^\alpha
V_{\omega-|\alpha|}w) = 0$ for $|\alpha| \leq \omega$, $c$ resp.\ $a$ 
are given by
\begin{align}
a^\alpha&=\scp{\tiltR{\omega}{w}}{x^\alpha V_{\omega-|\alpha|}w}, & 
c^\alpha&=\scp{\tiltR{\omega}{w}}{x^\alpha w}.
\label{eq:cundaaust}
\end{align}
They are related through:
\begin{align}
 a^\alpha&=c^\alpha\sum_{|\mu|\leq\omega-|\alpha|}
\frac{c^{\mu}}{\mu!}\D_{\mu}w^{-1}(0), &
 c^\alpha&=a^\alpha\sum_{|\mu|\leq\omega-|\alpha|}
\frac{a^{\mu}}{\mu!}\D_{\mu}w(0), &
1\leq|\alpha|\leq\omega, \notag \\
a^0&=\sum_{|\mu|\leq\omega}
\frac{c^{\mu}}{\mu!}\D_{\mu}w^{-1}(0), &
 c^0&=\sum_{|\mu|\leq\omega} \frac{a^{\mu}}{\mu!}\D_{\mu}w(0).
\label{eq:causa}
\end{align}
The equation for $a$ follows from the Leibnitz rule in
\eqref{def:tiltR}, while the equation for $c$ is derived from
\eqref{eq:cundaaust}.

In quantum field theory the coefficients $a$ are called counter terms.
They are not arbitrary. They have to be chosen in such a way that
Lorentz covariance of $^{0}t$ is preserved in the extension.  This
follows in the next sections. The remaining freedom is further
restricted by discrete symmetries like permutation symmetry or $C,P$
and $T$ symmetries. At the end gauge invariance or renormalization
constraints will fix the extension uniquely. But up to now there is no
local prescription of the latter.
\section{The $G$-covariant extension}
\label{sec:g}
We will first define the notion of a $G$-covariant distribution.  So
let $G$ be a linear transformation group on \Rd\ i.e.  $x\mapsto gx$,
$g\in G$. Then
\begin{equation}
        x^{\alpha}\mapsto{g^{\alpha}}_{\beta}x^{\beta}=(gx)^{\alpha}
        \label{}
\end{equation}
denotes the corresponding tensor representation.  $G$ acts on
functions in the following way:
 \begin{equation}
 (g\varphi)(x)\doteq\varphi(g^{-1}x),
 \label{def:gphi}
 \end{equation}
so that \Dcal\ is made a $G$-module.  We further have
\begin{align}
g(\varphi\psi)&=(g\varphi)(g\psi),\label{eq:gprod} \\
x^{\alpha}\D_{\alpha}(g^{-1}\varphi)
&=(gx)^{\alpha}g^{-1}(\D_{\alpha}\varphi), \label{eq:1} \\
x^{\alpha}\D_{\alpha}(g^{-1}\varphi)(0)
&=(gx)^{\alpha}\D_{\alpha}\varphi(0). \label{eq:2}
\end{align}
Now assume we have a distribution ${^0t}\in\Dcal'(\Rdon)$ that transforms 
covariantly under the Group $G$ as a density, i.e.
\begin{equation}
{^0t}(gx)|\det g|=D(g){^0t}(x),
\label{eq:tcov}
\end{equation}
where $D$ is the corresponding representation.  That means:
\begin{equation}
\scp{{^0t}}{g\psi}=\scp{D(g){^0t}}{\psi}\doteq D(g)\scp{{^0t}}{\psi}. 
\label{def:tcov}
\end{equation}
We will now investigate what happens to the covariance in the 
extension process. We compute:
\begin{align}
D(g)&\scp{\tR{\omega}{w}}{g^{-1}\varphi} 
-\scp{\tR{\omega}{w}}{\varphi} 
\label{eq:cov1} \\
&=D(g)\scp{{^0t}}{\W{\omega}{w}g^{-1}\varphi} 
-\scp{{^0t}}{\W{\omega}{w}\varphi} 
\label{eq:cov2} \\
&\stackrel{\eqref{eq:varW}}{=}
D(g)\scp{{^0t}}{g^{-1}\varphi-\sum_{|\alpha|\leq\omega} 
\frac{x^{\alpha}}{\alpha!}V_{\omega-|\alpha|}w\,\D_{\alpha}(g^{-1}\varphi)(0)} 
-\scp{{^0t}}{\W{\omega}{w}\varphi} \label{eq:cov3} \\
&\stackrel{(\ref{eq:gprod},\ref{eq:2})}{=}
D(g)\scp{{^0t}}{g^{-1}\biggl(\varphi- \sum_{|\alpha|\leq\omega} 
\frac{x^{\alpha}}{\alpha!}\left(gV_{\omega-|\alpha|}w\right)
\D_{\alpha}\varphi(0)\biggr)} 
-\scp{{^0t}}{\W{\omega}{w}\varphi} 
\label{eq:cov4} \displaybreak[0]\\
&\stackrel{(\ref{def:tcov})}{=}
\sum_{|\alpha|\leq\omega} 
\scp{{^0t}}{x^{\alpha}(\eins-g)(V_{\omega-|\alpha|}w)} 
\frac{\D_{\alpha}\varphi(0)}{\alpha!} \label{eq:cov8} \\
&\doteq\sum_{|\alpha|\leq\omega}b^{\alpha}(g) 
\frac{\D_{\alpha}\varphi(0)}{\alpha!}.
\label{def:balpha}
\end{align}
Then (\ref{def:balpha}) defines a map from $G$ to a finite dimensional
complex vectorspace. Now we follow \cite{prep:stora-pop},
\cite{bk:scharf}[chapter 4.5]: Applying two transformations
\begin{align}
b^{\alpha}(g_{1}g_{2})
&=\scp{{^0t}}{x^{\alpha}(\eins-g_{1}g_{2})(V_{\omega-|\alpha|}w)} \\
&=\scp{{^0t}}{x^{\alpha}\left((\eins-g_{1})
+g_{1}(\eins-g_{2})\right)(V_{\omega-|\alpha|}w)} \\
&=b^{\alpha}(g_{1})+|\det g_{1}|\scp{{^0t}(g_{1}x)} 
{(g_{1}x)^{\alpha}(\eins-g_{2})(V_{\omega-|\alpha|}w)},
\end{align} 
and omitting the indices we see
$b(g_{1}g_{2})=b(g_{1})+D(g_{1})g_{1}b(g_{2}), $ which is a 1-cocycle
for $ b(g) $.  Its trivial solutions are the 1-coboundaries
\begin{equation}
        b(g)=(\eins-D(g)g)a,
\label{def:cobound}
\end{equation}
and these are the only ones if the first cohomology group of $G$ is
zero. In that case we can restore $G$-covariance by adding the
following counter terms:
\begin{equation}
\scp{\tR{\omega}{w}^{G-\mathrm{cov}}}{\varphi} 
\doteq\scp{\tR{\omega}{w}}{\varphi} 
+\sum_{|\alpha|\leq\omega}
\frac{1}{\alpha!}a^{\alpha}(w)\D_{\alpha}\varphi(0).
 \label{def:tRGcov}
 \end{equation} 
The task is to determine $a$ from (\ref{def:cobound}) and (\ref{eq:cov8}):
\begin{equation}
\scp{{^0t}}{x^{\alpha}(\eins-g)(V_{\omega-|\alpha|}w)}
=\bigl[(\eins-D(g)g)a\bigr]^\alpha
\label{def:a}
\end{equation}

\section{Tensorial Lorentz covariance}
\label{sec:tensor}
The first cohomology group of $\Lcal_{+}^{\uparrow}$ vanishes
\cite{bk:scharf}[chapter 4.5 and references there]. We determine $ a $
from the last equation. The most simple solution appears in the case
of Lorentz invarinance in one coordinate. This situation was
completely analyzed in \cite{prep:wir} for $\D_{\alpha}w(0) =
\delta_{\alpha}^{0}$.  The following two subsections generalize the
results to arbitrary $w,\ w(0)\not=0$.

\subsection{Lorentz invariance in \Rv}
\label{subsec:Linv4d}
If we expand the index $\alpha$ into Lorentz indices $\mu_1, \dots, 
\mu_n$, \eqref{def:a} is symmetric in $\mu_1, \dots, \mu_n$ and 
therefore $a$ will be, too. We just state our result from 
\cite{prep:wir} which is modified through the generalization for the 
choice of $w$:
\begin{multline}
  a^{(\mu _1 \ldots \mu_n)} = \frac{ (n-1)!!}{(n+2)!!}
  \sum_{s=0}^{\left[\frac{n-1}{2} \right] }
  \frac{(n-2s)!!}{(n-2s-1)!!} \eta^{(\mu_1 \mu _2} 
  \ldots \eta^{\mu_{2s-1}\mu _{2s}} \times \\ 
  \times \left\langle 
  {^0t}, (x^2)^s x ^{\mu _{2s+1}} \ldots x ^{\mu_{n-1}} 
  \left(x^2\D^{\mu_{n})}-x^{\mu_{n})}x^{\beta}\D_{\beta}\right)
  V_{\omega-n}w\right\rangle,
\label{eq:amun}
\end{multline}
if we choose the fully contracted part of $ a $ to be zero in case 
of $ n $ being even. We used the notation
\begin{align*}
b^{(\mu_{1}\dots\mu_{n})}&=\frac{1}{n!}\sum_{\pi\in S_{n}} 
b^{\mu_{\pi(1)}\dots\mu_{\pi(n)}}, &
b^{[\mu_{1}\dots\mu_{n}]}&=\frac{1}{n!}\sum_{\pi\in S_{n}} 
\sgn(\pi)b^{\mu_{\pi(1)}\dots\mu_{\pi(n)}},
\end{align*}
for the totally symmetric resp.\ antisymmetric part of a tensor.


\subsection{Dependence on $w$}
\label{subsubsec:scale} 
Performing a functional derivation of the Lorentz invariant extension
with respect to $w$, only Lorentz invariant counter terms appear.
\begin{definition}
The functional derivation is given by:
\begin{equation*}
\scp{\frac{\delta}{\delta g}F(g)}{\psi}\doteq
\left.\frac{\dif}{\dif\lambda}F(g+\lambda\psi)\right\vert_{\lambda=0}. 
\end{equation*}
\end{definition}
This definition implies the following functional derivatives:
\begin{align}
\scp{\frac{\delta}{\delta w}\tR{\omega}{w}(\varphi)}{\psi}
&=-\sum_{|\alpha|\leq\omega}\frac{1}{\alpha!}
\scp{\tR{\omega}{w}}{x^\alpha\psi}
\D_\alpha\left(\varphi w^{-1}\right)(0),\label{eq:derivtR} \\
\scp{\frac{\delta}{\delta w}\scp{S}{V_\omega w}}{\psi}
&=\sum_{|\alpha|\leq\omega}\frac{1}{\alpha!}
\scp{S}{\W{\omega}{w}(x^\alpha\psi)}
\D_\alpha w^{-1} (0), \label{eq:derivS}
\end{align}
for any distribution $S$. 
\begin{proof} 
  We show how to derive the first relation. Inserting the definition we 
  find:
\begin{align*}
\left.\frac{\dif}{\dif\lambda}
\tR{\omega}{w+\lambda\psi}(\varphi)\right\vert_{\lambda=0}
&=\scp{^0t}{-\psi\sum_{|\alpha|\leq\omega}
\frac{x^\alpha}{\alpha!}\D_\alpha\left(\varphi w^{-1}\right)(0)
+w\sum_{|\alpha|\leq\omega}\frac{x^\alpha}{\alpha!} 
\D_\alpha\left(\varphi\psi w^{-2}\right)(0)}, \\
\intertext{using Leibnitz rule and rearranging the summation in the
  second term,}
&\begin{aligned}
=\sum_{|\alpha|\leq\omega}\frac{1}{\alpha!}
\Biggl\langle{^0t},-&\psi x^\alpha\D_\alpha\left(\varphi w^{-1}\right)(0) \\
+&w x^\alpha\D_\alpha\left(\varphi w^{-1}\right)(0)
\sum_{|\nu|\leq\omega-|\alpha|}\frac{x^\nu}{\nu!}
\D_\nu\left(\psi w^{-1}\right)(0) \Biggr\rangle
\end{aligned} \\
&=-\sum_{|\alpha|\leq\omega}\frac{1}{\alpha!}
\scp{^0t}{x^\alpha\W{\omega-|\alpha|}{w}\psi}
\D_\alpha\left(\varphi w^{-1}\right)(0) \\
&=-\sum_{|\alpha|\leq\omega}\frac{1}{\alpha!}
\scp{\tR{\omega}{w}}{x^\alpha\psi}\D_\alpha\left(\varphi w^{-1}\right)(0),
\end{align*}
where we used the relation $x^\alpha\W{\omega-|\alpha|}{w}\varphi = 
\W{\omega}{w}(x^\alpha\varphi)$ on the last line. The second equation 
follows from a similar calculation.
\end{proof}
We calculate the dependce of $a$ on $w$. With \eqref{eq:derivS} we
get:
\begin{multline*}
\frac{\delta}{\delta w} a^{(\mu_1\dots\mu_n)}(w)
= \frac{(n-1)!!}{(n+2)!!}
  \sum_{s=0}^{\left[\frac{n-1}{2} \right] }
  \frac{(n-2s)!!}{(n-1-2s)!!}  \eta^{(\mu_1\mu_2} \dots
  \eta^{\mu _{2s-1} \mu _{2s}} \times \\ 
\times\sum_{|\beta|\leq\omega-n}\frac{1}{\beta!}
  \scp{^0t}{(x^2)^s x^{\mu_{2s+1}} \dots x ^{\mu_{n-1}} 
  \left(x^2\D^{\mu _{n})} 
  -x^{\mu_{n})}x^{\beta}\D_{\beta}\right)\W{\omega-n}{w}(x^\beta\psi)}
\D_\beta w^{-1}(0).
\label{eq:damun}
\end{multline*}
To condense the notation we again use $\beta$ as a multiindex. Since 
$\W{\omega-n}{w}(x^\beta\psi)$ is sufficient regular, we can put the 
$x$'s and derivatives on the left and the same calculation like in 
\cite{prep:wir} applies. The result is
\begin{multline*}
  \scp{\frac{\delta}{\delta w}a^{\mu_{1}\cdots\mu_{n}}(w)}{\psi}=
  \sum_{|\beta|\leq\omega-n}\frac{\D_\beta w^{-1}(0)}{\beta!}
\Biggl[
\scp{\tR{\omega}{w}}{x^{\mu_{1}}\cdots x^{\mu_{n}}x^\beta\psi}
 + \\
-\begin{cases}
 0,&n\text{ odd}, \\
 \frac{2(n-1)!!}{(n+2)!!}  
 \scp{\tR{\omega}{w}}{(x^2)^\frac{n}{2}x^\beta\psi}
 \eta^{(\mu_1\mu_2}\cdots\eta^{\mu_{n-1}\mu_n)}, 
 &n\text{ even.}
\end{cases}
\Biggr]
\end{multline*}
Using this result and \eqref{eq:derivtR} we find:
\begin{align*}
  \scp{\frac{\delta}{\delta w}\scp{\tRli{\omega}{w}}{\phi}}{\psi}
  &=-\sum_{\substack{n=0\\ n\text{ even}}}^{\omega}\frac{d_{n}}{n!}
  \square^{\frac{n}{2}}\phi(0), \\
  d_{n}&\doteq
  \frac{2(n-1)!!}{(n+2)!!}\sum_{|\beta|\leq\omega-n}\frac{1}{\beta!}
\scp{\tR{\omega}{w}}{(x^{2})^{\frac{n}{2}}x^\beta\psi}
\D_\beta w^{-1}(0), 
\end{align*}
where we set $ d_{0}=1 $.


\subsection{General Lorentz covariance}
If the distribution ${^{0}t}$ depends on more than one variable, 
${^{0}t}x^{\alpha}$ will not be symmetric in all Lorentz indices in general.
Since $x^{\alpha}$ transforms like a tensor, it is natural to 
generalize the discussion to the case, where ${^{0}t}$ transforms like a 
tensor, too. Assume rank$( {^0t} )=r$, then $D(g)g$ is the tensor 
representation of rank $p=r+n,n=|\alpha|$, in (\ref{def:a}). From now 
on we will omit the indices. So if $t\in\Dcal(\Rvmon)$, we denote by 
\xbar\ -- formerly $x^{\alpha}$ -- a tensor of rank $n$ built of 
$x_{1},\dots,x_{m}$. 

To solve \eqref{def:a} we proceed like in 
\cite{prep:wir}. Since the equation holds for all $g$ we will solve 
for $a$ by using Lorentz transformations in the infinitesimal 
neighbourhood of $\eins$. If we take 
$\theta_{\alpha\beta}=\theta_{[\alpha\beta]} $ as six coordinates these transformations read:
\begin{equation}
g \approx\eins+\frac{1}{2}\theta_{\alpha\beta}l^{\alpha\beta},
\label{eq:gbyone}
\end{equation}
with the generators
\begin{equation}
{(l^{\alpha\beta})^{\mu}}_{\nu}
=\eta^{\alpha\mu}\delta^{\beta}_{\nu}-\eta^{\beta\mu}\delta^{\alpha}_{\nu}.
\label{def:generator}
\end{equation}
Then, for an infinitesimal transformation one
finds from \eqref{def:a}:
\begin{gather}
B^{\alpha\beta}\doteq 
2\biggl\langle{^0t},\xbar\sum_{j=1}^{m}
x^{[\alpha}_j\D^{\beta]}_j(V_{\omega-n}w)\biggr\rangle 
=(l^{\alpha\beta}\otimes\dots\otimes\eins+\dots 
+\eins\otimes\dots\otimes l^{\alpha\beta})a, 
\label{eq:atensor}
\end{gather}
$\alpha,\beta$ being Lorentz four-indices. In \cite{prep:wir}
our ability to solve that equation heavily relied on the given 
symmetry, which is in general absent here. Nevertheless we can find 
an inductive construction for $a$, corresponding to equation 
(29) in \cite{prep:wir}. 

We build one Casimir operator on the r.h.s. (the other one is always 
zero, since we are in a $(1/2,1/2)^{\otimes p}$ representation).
\subsubsection*{The case $p=1$} 
Just to remind that $p$ is the rank of $\xbar t$, this occurs if
either $t$ is a vector and $\xbar=1, (n=0)$, or $t$ is a scalar and
$\xbar=x_{1},\dots,x_{m}$.  (\ref{eq:atensor}) gives:
\begin{equation}
\frac{1}{2}l_{\alpha\beta}B^{\alpha\beta}
=\frac{1}{2}l_{\alpha\beta}l^{\alpha\beta}a 
=-3\eins a,
\label{eq:cas1}
\end{equation}
since the Casimir operator is diagonal in the irreducible $(1/2,1/2)$ 
representation. 

\subsubsection*{The case $p=2$}
We get
\begin{equation}
\frac{1}{2}(l_{\alpha\beta}\otimes\eins
+\eins\otimes l_{\alpha\beta})B^{\alpha\beta}
=(-6\eins+l_{\alpha\beta}\otimes l^{\alpha\beta})a.
\label{eq:acas2}
\end{equation}
Since $a$ is a tensor of rank 2, let us introduce the projector onto the
symmetric resp. antisymmetric part and the trace:
\begin{align}
{P_{S}^{\mu\nu}}_{\rho\sigma}
&=\frac{1}{2}(\delta^{\mu}_{\rho}\delta^{\nu}_{\sigma}
+\delta^{\nu}_{\rho}\delta^{\mu}_{\sigma}), &
{P_{A}^{\mu\nu}}_{\rho\sigma}
&=\frac{1}{2}(\delta^{\mu}_{\rho}\delta^{\nu}_{\sigma}
-\delta^{\nu}_{\rho}\delta^{\mu}_{\sigma}), \label{def:P} &
{P_\eta^{\mu\nu}}_{\rho\sigma}
&=\frac{1}{4}\eta^{\mu\nu}\eta_{\rho\sigma}, \\
P^2&=P, &
P_S+P_A&=\eins, &
P_S-P_A&=\tau, 
\end{align}
where $\tau$ denotes the permutation of the two indices. 
Using (\ref{def:generator}), we find
\begin{equation}
\frac{1}{2}l_{\alpha\beta}\otimes l^{\alpha\beta}=4P_{\eta}-\tau.
\label{eq:genten}
\end{equation}
Now we insert (\ref{eq:genten}) into (\ref{eq:acas2}). The trace part
will be set to zero again. Acting with $P_A$ and $P_S$ on the
resulting equation gives us two equations for the antisymmetric and
symmetric part respectively. This yields:
\begin{equation}
a=-\frac{1}{16}(P_S+2P_A)(l_{\alpha\beta}\otimes\eins
+\eins\otimes l_{\alpha\beta})B^{\alpha\beta}.
\label{eq:a2}
\end{equation}

\subsubsection*{Inductive assumption}
Now we turn back to equation \eqref{def:a}. We note that any
contraction commutes with the (group) action on the rhs. Hence, if we
contract \eqref{eq:atensor}, we find on the rhs:
\begin{multline*}
\eta_{ij}(l^{\alpha\beta}\otimes\dots\otimes\eins+\dots 
+\eins\otimes\dots\otimes l^{\alpha\beta})a= \\
(l^{\alpha\beta}\otimes\dots\otimes\eins+\dots
+\widehat{i}+\dots+\widehat{j}+\dots
+\eins\otimes\dots\otimes l^{\alpha\beta})(\eta_{ij}a),
\end{multline*}
where $i,j$ denote the positions of the corresponding indices, and the
$\widehat{\text{hat}}$ means omission. Therefore the rank of
\eqref{eq:atensor} is reduced by two and we can proceed inductively.
With the cases $p=1, p=2$ solved, we assume that all possible
contractions of $a$ are known.

\subsubsection*{Induction step}
Multiplying (\ref{eq:atensor}) with the generator and contracting the
indices yields:
\begin{multline}
\biggl(3p\eins+2\sum_{\tau\in S_p}\tau\biggr)a 
=-\frac{1}{2}(l_{\alpha\beta}\otimes\dots\otimes\eins+\dots 
+\eins\otimes\dots\otimes l_{\alpha\beta})B^{\alpha\beta}
+8\sum_{i<j\leq p}{P_\eta}_{ij}a.
\label{eq:acasp}
\end{multline}
The transposition $\tau$ acts on $a$ by permutation of the
corresponding indices. For a general $\pi\in S_p$ the action on $a$ is
given by: $\pi
a^{\mu_1\dots\mu_p}=a^{\mu_{\pi^{-1}(1)}\dots\mu_{\pi^{-1}(p)}}$. In
order to solve this equation we consider the representation of the
symmetric groups. We give a brief summary of all necessary ingredients
in appendix~\ref{app:permrep}. So let $k_\tau\doteq\sum_{\tau\in
  S_p}\tau$ be the sum of all transpositions of $S_p$.  Then $k_\tau$ is
in the center of the group algebra $\Acal_{S_p}$.  It can be
decomposed into the idempotents $e_{(m)}$ that generate the irreducible
representations of $S_p$ in $\Acal_{S_p}$.
\begin{equation}
k_\tau=h_\tau\sum_{(m)}\frac{1}{f_{(m)}}\chi_{(m)}(\tau)e_{(m)}.
\label{eq:decompktau}
\end{equation}
The sum runs over all partitions $(m)=(m_1,\dots,m_r), \sum_{i=1}^r
m_i=p, m_1\geq m_2\geq\dots\geq m_r$ and $h_\tau=\frac{1}{2} p(p-1)$
is the number of transpositions in $S_p$. $\chi_{(m)}$ is the
character of $\tau$ in the representation generated by $e_{(m)}$ which
is of dimension $f_{(m)}$. We use \eqref{eq:decompktau}, the
ortogonality relation $e_{(m)}e_{(m')} =\delta_{(m)(m')}$ and the
completeness $\sum_{(m)}e_{(m)}=\eins$ in (\ref{eq:acasp}). The
expression in brackets on the l.h.s may be orthogonal to some
$e_{(m)}$. The corresponding $e_{(m)}a$ contribution will be any
combinations of $\eta$'s and $\epsilon$'s --$\epsilon$ being the
totally antisymmetric tensor in four dimensions -- transforming
correctly and thus can be set to zero. We arrive at
\begin{multline}
a=\sum_{\substack{(m)\\c(m)\not=0}}\frac{e_{(m)}}{c(m)}
\left(
-\frac{1}{2}(l_{\alpha\beta}\otimes\dots\otimes\eins+\dots 
+\eins\otimes\dots\otimes l_{\alpha\beta})B^{\alpha\beta}
+8\sum_{i<j\leq p}{P_\eta}_{ij}a
\right), \\
c(m)\doteq 3p+p(p-1)\frac{\chi_{(m)}(\tau)}{f_{(m)}}
=3p+\sum_{i=1}^r\left(
b_i^{(m)}(b_i^{(m)}+1)-a_i^{(m)}(a_i^{(m)}+1)
\right),
\label{eq:atenrec}
\end{multline}
with $a=(a_1,\dots,a_r), b=(b_1,\dots,b_r)$ denoting the
characteristics of the frame $(m)$, see appendix~\ref{app:permrep}. 
Let us take $p=4$ as an example:
\[
\begin{array}{cccc}
\text{idempotent} & \text{Young frame} & \text{dimension} & 
\text{character} \\
& & & \\
e_{(4)}       & \yng(4)       & f_{(4)}=1       & \chi_{(4)}(\tau)=1  \\
& & & \\
e_{(3,1)}     & \yng(3,1)     & f_{(3,1)}=3     & \chi_{(3,1)}(\tau)=1  \\
& & & \\
e_{(2,2)}     & \yng(2,2)     & f_{(2,2)}=2     & \chi_{(2,2)}(\tau)=0  \\
& & & \\
e_{(2,1,1)}   & \yng(2,1,1)   & f_{(2,1,1)}=3   & \chi_{(2,1,1)}(\tau)=-1 \\
& & & \\
e_{(1,1,1,1)} & \yng(1,1,1,1) & f_{(1,1,1,1)}=1 & \chi_{(1,1,1,1)}(\tau)=-1
\end{array}
\]
We find for (\ref{eq:acasp})
\begin{equation}
a=\frac{1}{48}(2e_{(4)} + 3e_{(3,1)} +4e_{(2,2)} +6e_{(2,1,1)})\times
\text{r.h.s}(\ref{eq:acasp}).
\end{equation}
We see that no $e_{(1,1,1,1)}$ appears in that equation. It corresponds to the one dimensional
$sgn$-representation of $S_4$, so $e_4a\propto\epsilon$.

\section{Spinorial Lorentz covariance}
\label{sec:spinor}
In this section we follow \cite{bk:sexl-urb}.  The most general
representation of $\Lcal_{+}^{\uparrow}$ can be built of tensor
products of \SLC\ and $\overline{\SLC}$ and direct sums of these.  A
two component spinor $\Psi$ transforms according to
\begin{equation}
\Psi^{A}={g^{A}}_{B}\Psi^{B},
\label{eq:spintrafo}
\end{equation}
where $g$ is a $2\times2$-matrix in the \SLC\ representation of
$\Lcal_{+}^{\uparrow}$.  For the complex conjugated representation we
use the dotted indices, i.e.
\begin{equation}
\overline{\Psi}^{\Xdot}={\overline{g}^{\Xdot}}_{\Ydot}\overline{\Psi}^{\Ydot},
\label{eq:ccspintrafo}
\end{equation}
with $ {\overline{g}^{\Xdot}}_{\Ydot}=\overline{{g^{X}}_{Y}} $ in the $ 
\overline{\SLC} $ representation.  The indices are lowered and raised 
with the $ \epsilon $-tensor.
\begin{gather}
\epsilon_{AB}=\overline{\epsilon}_{\Adot\Bdot}
\doteq\epsilon_{\Adot\Bdot}, \label{eq:eps} \\
\epsilon^{AB}\epsilon_{AC}
=\epsilon^{BA}\epsilon_{CA}
=\delta^{B}_{C}.
\end{gather}
We define the Van-der-Waerden symbols with the help of the Pauli 
matrices $\sigma_{\mu}$ and ${\widetilde{\sigma}}_{\mu}\doteq\sigma^{\mu}$:
\begin{align}
{\sigma_{\mu}}^{A\Xdot}&\doteq\frac{1}{\sqrt{2}}(\sigma_{\mu})^{AX}, &
{\sigma_{\mu}}_{A\Xdot}&\doteq\frac{1}{\sqrt{2}}
({{\widetilde{\sigma}}_{\mu}}^{T})_{AX}.
\label{def:vdw}
\end{align}
They satisfy the following relations
\begin{align}
{\sigma_{\mu}}^{A\Xdot}{\sigma_{\nu}}_{A\Xdot}
&=\eta_{\mu\nu} &
{\sigma_{\mu}}_{A\Xdot}{\sigma^{\mu}}_{B\Ydot}
&=\epsilon_{AB}\epsilon_{\Xdot\Ydot}
\label{eq:vdw}
\end{align}
With the help of these we can build the infinitesimal spinor
transformations
\begin{equation}
g\approx\eins+\frac{1}{2}\theta_{\alpha\beta}S^{\alpha\beta},
\label{eq:spininf}
\end{equation}
with the generators
\begin{equation}
{(S^{\alpha\beta})^{A}}_{B}
={\sigma^{[\alpha}}^{A\Xdot}{\sigma^{\beta]}}_{B\Xdot}.
\label{eq:spingen}
\end{equation}
Note that the $ \sigma $'s are hermitian: $
\overline{{\sigma_{\mu}}^{A\Xdot}}={\sigma_{\mu}}^{X\Adot} $.  Again
we define the projectors for the tensor product. But we have only two
irreducible parts:
\begin{align}
{{P_S}^{AB}}_{CD}&=\frac{1}{2}
(\delta^A_C\delta^B_D+\delta^A_D\delta^B_C), &
{{P_\epsilon}^{AB}}_{CD}&=\frac{1}{2}
\epsilon^{AB}\epsilon_{CD}, \\
P^2=P, & 
P_S+P_\epsilon&=\eins.
\end{align}
We get the following identities:
\begin{align}
S^{\alpha\beta}S_{\alpha\beta}
&=\overline{S}^{\alpha\beta}\overline{S}_{\alpha\beta}
=-3\eins, \label{eq:sgen1} \\
S^{\alpha\beta}\otimes S_{\alpha\beta}
&=\overline{S}^{\alpha\beta}\otimes\overline{S}_{\alpha\beta}
=4P_\epsilon-\eins \label{eq:sgen2}, \\
S^{\alpha\beta}\otimes\overline{S}_{\alpha\beta}
&=\overline{S}^{\alpha\beta}\otimes S_{\alpha\beta}
=0. \label{eq:sgen3}
\end{align}

In order to have (\ref{def:a}) in a pure spinor representation we have
to decompose the tensor $\xbar$ into spinor indices according to
\begin{equation}
x^{A\Xdot}\doteq x^\mu{\sigma_{\mu}}^{A\Xdot}.
\end{equation}
Assume $t\widetilde{x}$ transforms under the $u$-fold tensor product of
\SLC\ times the $v$-fold tensor product of $\overline{\SLC}$ then,
for infinitesimal transformations, (\ref{def:a}) yields:
\begin{equation}
B^{\alpha\beta}
=(S^{\alpha\beta}\otimes\dots\otimes\eins+\dots
+\eins\otimes\dots\otimes\overline{S}^{\alpha\beta})a,
\end{equation}
with $B^{\alpha\beta}$ from equation \eqref{eq:atensor} in the
corresponding spinor representation.  The sum consists of $u$ summands
with one $S^{\alpha\beta}$ and $v$ summands with one
$\overline{S}^{\alpha\beta}$ with $u,v>n$.  Multiplying again with the
generator and contracting the indices gives twice the Casimir on the
r.h.s. Inserting (\ref{eq:sgen1}-\ref{eq:sgen3}) yields:
\begin{multline}
(S_{\alpha\beta}\otimes\dots\otimes\eins+\dots
+\eins\otimes\dots\otimes\overline{S}_{\alpha\beta})B^{\alpha\beta} \\
=\biggl(
-3(u+v)\eins
+2\sum_{1\leq i<j\leq u}(4P_{\epsilon_{ij}}-\eins) +
2\sum_{1\leq i<j\leq v}(4P_{\overline{\epsilon}_{ij}}-\eins) 
\biggr)a.
\end{multline}
The sum over $u$ runs over $\frac{u}{2}(u-1)$ possibilities and
similar for $v$, so we find the induction:
\begin{multline}
a=\frac{1}{u(u+2)+v(v+2)}\Biggl[
-(S_{\alpha\beta}\otimes\dots\otimes\eins+\dots
+\eins\otimes\dots\otimes\overline{S}_{\alpha\beta})B^{\alpha\beta}+
\\
8\biggl(
\sum_{1\leq i<j\leq u}P_{\epsilon_{ij}} 
+\sum_{1\leq i<j\leq v}P_{\overline{\epsilon}_{ij}} 
\biggr)a \Biggr].
\label{eq:spinind}
\end{multline}
It already contains the induction start for $a^{(AB)}, a^{(XY)}$ and 
$a^{A\Xdot}$. 

\section{General covariant BPHZ subtraction}
\label{sec:bphz}
In this section we shrink the distribution space to $\Scal'$ since we are 
dealing with Fourier transformation. Let $x,q,p\in\Rvm$. BPHZ subtraction 
at momentum $q$ corresponds to using $w=e^{iq\cdot}$ in the 
EG subtraction \cite{pap:prange1}.
\begin{align}
\widehat{\tR{\omega}{e^{iq\cdot}}}(p)
&\doteq\scp{\tR{\omega}{e^{iq\cdot}}} {e^{ip\cdot}} \\
&=\scp{^{0}t}{e^{ip\cdot}-\sum_{|\alpha|\leq\omega} 
\frac{(p-q)^\alpha}{\alpha!}\D^q_{\alpha}e^{iq\cdot}}.
\label{eq:BPHZ}
\end{align}
It is normalized at the subtraction point $q$, i.e.:
$\D_{\alpha}\widehat{\tR{\omega}{e^{iq\cdot}}}(q) = 0,\ |\alpha| \leq
\omega$.  This is always possible for $q$ totally spacelike,
$(\sum_{j\in I}q_{j})^2<0, \forall I\subset\{1,\dots,m\}$
\cite{pap:ep-gl,priv:duetsch}. In massive theories one can put $q=0$
and has the usual subtraction at zero momentum which preserves
covariance.  But this leads to infrared divergencies in the massless
case. There we can use the results from above to construct a covariant
BPHZ subtraction for momentum $q$ by adding
$\sum_{|\alpha|\leq\omega} \frac{i^{|\alpha|}}{\alpha!} a^\alpha
p_\alpha$ to (\ref{eq:BPHZ}), according to equation \eqref{def:tRGcov}.
For $|\beta|\geq\omega+1$, ${^0t}x^\beta$ is a well defined distribution
on $\Scal$ and so is $\D_\beta\widehat{^0t}$.

\subsection{Lorentz invariance on \Rv} 
We have
\begin{align}
V_k e^{iq\cdot}&=e^{iq\cdot}\sum_{m=0}^k\frac{1}{m!}(-iqx)^m, & 
\D_\sigma V_k e^{iq\cdot} &=iq_\sigma e^{iq\cdot}\frac{1}{k!}(-iqx)^k.
\end{align}
Inserting this into (\ref{eq:amun}) we find:
\begin{multline}
a^{(\mu _1\dots\mu_n)}=\frac{i^n(-)^{\omega+1}}{(\omega-n)!}
\frac{ (n-1)!!}{(n+2)!!}
q_{\sigma_1}\dots q_{\sigma_{\omega-n}} 
\sum_{s=0}^{\left[\frac{n-1}{2} \right] }
\frac{(n-2s)!!}{(n-2s-1)!!}
\left(q_\rho\D^\rho\D^{(\mu_1}-q^{(\mu_1}\square\right)\times \\ 
\times \eta^{\mu_2\mu_3}\dots\eta^{\mu_{2s}\mu_{2s+1}} 
\D^{\mu_{2s+2}}\dots\D^{\mu_n)}\square^s
\D^{\sigma_1}\dots\D^{\sigma_{\omega-n}}\widehat{{^0t}}(q).
\label{eq:amunq}
\end{multline}

\subsubsection*{Example}
Take the setting sun in massless scalar field theory: 
$^{0}t=\frac{i^{3}}{6}D_{F}^{3} \Rightarrow \omega=2$.
\begin{align*}
a^\mu&=-\frac{i}{3}(q_\sigma q_\rho \D^\rho \D^\sigma \D^\mu
-q^\mu q_\sigma\D^\sigma\square)\widehat{^{0}t}(q), \\
a^{\mu\nu}&=\frac{1}{4}(q_\rho \D^\rho \D^\mu \D^\nu 
-q^{(\mu}\D^{\nu)}\square)\widehat{^{0}t}(q),
\end{align*}
and adding $ip_{\mu}a^{\mu} - \frac{1}{2}p_{\mu}p_{\nu}a^{\mu\nu}$ 
restores Lorentz invariance of the setting sun graph subtracted at $q$.
 
\subsection{General induction}
We only have to evaluate $B^{\alpha\beta}$ with $w=e^{iq\cdot}$ and
plug the result into the induction formulas \eqref{eq:spinind} and
\eqref{eq:atenrec}.
\begin{equation}
B^{\alpha\beta}
=2i^{n}(-)^{\omega+1}\sum_{j=1}^{m}\sum_{|\gamma|=\omega-n}
\frac{q^{\gamma}}{\gamma!}q_{j}^{[\alpha}\D_{j}^{\beta]}
\D_{\gamma}\widetilde{\D}\,\widehat{^{0}t}(q).
\end{equation}
Here, $q_{j}$ are the $m$ components of $q$ hence $\gamma$ is a $4m$
index and $\alpha,\beta$ are four indices. The tensor (spinor)
structure of $\widetilde{\D}$ is given by $\widetilde{x}$ in
\eqref{eq:atensor}.
 
\section{Summary and Outlook}
The subtraction procedure in EG renormalization makes use of an
auxiliary (test) function and hence breaks Lorentz covariance, since
no Lorentz invariant test function exists. But this symmetry can be
restored by an appropriate choice of counterterms. We give an explicit
formula for their calculation in lowest order and an inductive one for
higher orders. Using the close relationship to BPHZ subtraction this
directly translates into a covariant subtraction at totally spacelike
momentum.

We expect our solution to be useful for all calculations for which the
central solution ($w=1$, see \cite{bk:scharf}) does not exist, namely
all theories that contain loops of only massless particles.

\section{Acknowledgment}
I would like to thank Klaus Bresser and Gudrun Pinter for our short
but efficient teamwork and K. Fredenhagen for permanent support.

\clearpage


\appendix


\section{Representation of the symmetric groups}
\label{app:permrep}
Everything in this brief appendix should be found in any book about
representation theory of finite groups. We refer to
\cite{bk:simon,bk:boerner,bk:fulton}.  The group algebra
$\Acal_{S_p}$ consists of elements
\begin{align}
a&=\sum_{g\in S_p}\alpha(g)\cdot g,&
b&=\sum_{g\in S_p}\beta(g)\cdot g,
\end{align}
where $\alpha,\beta$ are arbitrary complex numbers. The sum of two 
elements
is naturally given by the summation in $\menge{C}$ and the product is
defined through the following convolution:
\begin{align}
ab&\doteq\sum_{g_1,g_2}\alpha(g_1)\beta(g_2)\cdot g_1 g_2
=\sum_g\gamma(g)\cdot g \text{ with} \\
\gamma(g)&\doteq\sum_{g_1 g_2=g}\alpha(g_1)\beta(g_2)
=\sum_{g_1}\alpha(g_1)\beta({g_1}^{-1}g)
=\sum_{g_2}\alpha(g{g_2}^{-1})\beta(g_2).
\end{align}
The group algebra is the direct sum of simple twosided ideals:
\begin{equation}
\Acal_{S_p}=I_1\oplus\dots\oplus I_k,
\end{equation}
and $k$ is the number of partitions of $p$. Every ideal $I_j$ contains
$f_j$ equivalent irreducible representations of $S_p$. $I_j$ is
generated by an idempotent $e_j\in\Acal_{S_p}$:
\begin{align}
I_j&=\Acal_{S_p}e_j & {e_j}^2&=e_j.
\end{align}
These idempotents satisfy the following orthogonality and completeness
relations:
\begin{align}
e_j e_i &= \delta_{ji} & \sum_{j=1}^k e_j &=\eins.
\end{align}
The center of $\Acal_{S_p}$ consists of all elements $\sum_j^k\alpha_j
e_j, \alpha_j\in\menge{C}$.

Every permutation of $S_p$ can be uniquely (modulo order) written as a
product of disjoint cycles. Since two cycles are conjugated if and
only if their length is the same, the number of conjugacy classes is
equal to the number of partitions of $p$. Denoting the $j$'th
conjugacy class by $c_j$ we build the sum of all elements of one class
\begin{equation}
k_j\doteq\sum_{\pi\in c_j}\pi \in \Acal_{S_p}
\end{equation}
which is obviously in the center of $\Acal_{S_p}$, too. So we can
expand $k_i$ in the basis $e_j$:
\begin{equation}
k_i=h_i\sum_{j=1}^k\frac{1}{f_j}\chi_j(c_i)e_j,
\label{eq:decompki}
\end{equation}
where $\chi_j(c_i)$ is the character of the class $c_i$ in the
representation generated by $e_j$ and $h_i$ is the number of elements
of $c_i$. The dimension of that representation is equal to the
multiplicity $f_j$.

The constuction of the idempotents can be carried out via the 

\subsection*{Young taubleaux}
A sequence of integers $(m)=(m_1,\dots,m_r), m_1\geq m_2\geq\dots\geq
m_r$ with $\sum_{j=1}^r=p$ gives a partition of $p$. To every such
sequence we associate a diagram with
\[
\begin{array}{cl}
m_1\text{ boxes}&\yng(3)\dots\yng(1) \\[-.5pt]
m_2\text{ boxes}&\yng(2)\dots \\
\vdots &
\end{array}
\]
called a \emph{Young frame} $(m)$. Let us take $p=5$ as an example:
\[
\begin{array}{ccccccc}
\yng(5) & \yng(4,1) & \yng(3,2) & \yng(3,1,1) & \yng(2,2,1) &
\yng(2,1,1,1) & \yng(1,1,1,1,1) \\
(5) & (4,1) & (3,2) & (3,1,1) & (2,2,1) & (2,1,1,1) & (1,1,1,1,1) 
\end{array}
\]
An assignment of numbers $1,\dots,p$ into the boxes of a frame is
called a \emph{Young tableau}. Given a tableau $T$, we denote 
$(m)$ by $(m)(T)$. If the numbers in every row and in every column
increase the tableau is called \emph{standard}. The number of standard
tableaux for the frame $(m)$ is denoted by $f_{(m)}$. It is equal to the
dimension of the irreducible representation generated by the
idempotent $e_{(m)}$. We will now answer the question 

\subsubsection*{How to construct $e_{(m)}$}
Set
\begin{align*}
\Rcal(T)&\doteq
\{\pi\in S_p|\pi\text{ leaves each row of $T$ setwise fixed}\}, \\
\Ccal(T)&\doteq
\{\pi\in S_p|\pi\text{ leaves each column of $T$ setwise fixed}\},
\end{align*}
and build the following objects:
\begin{align}
P(T)&\doteq\sum_{p\in\Rcal(T)}p, & \label{def:PQ}
Q(T)&\doteq\sum_{q\in\Rcal(T)}\sgn(q) q,
\end{align}
then
\begin{equation}
e(T)\doteq\frac{f_{(m)}}{p!}P(T)Q(T) \label{def:eT}
\end{equation}
is a minimal projection in $\Acal_{S_p}$ (generates a minimal left
ideal). The central projection (generating the simple twosided ideal)
is given by
\begin{equation}
e_{(m)}\doteq\frac{f_{(m)}}{p!}\sum_{T|(m)(T)=(m)} e(T).
\label{def:eF}
\end{equation}

\subsection*{Example $p=3$}
The frame {\scriptsize\yng(3)} has only one standard tableau
{\scriptsize\young(123)}. All different tableaux in (\ref{def:eF})
lead to the same idempotent (\ref{def:eT}) which is just the sum of
all permutations.
\[
e_{(3)}=\frac{1}{6}
\bigl(\eins+(1\,2)+(1\,3)+(2\,3)+(1\,2\,3)+(1\,3\,2)\bigr).
\]
For the frame {\scriptsize\yng(1,1,1)} we only need the column permutations in
(\ref{def:eT}). We find 
\[
e_{(1,1,1)}=\frac{1}{6}
\bigl(\eins-(1\,2)-(1\,3)-(2\,3)+(1\,2\,3)+(1\,3\,2)\bigr).
\]
The frame {\scriptsize\yng(2,1)} has two standard tabelaux. For the
tableaux {\scriptsize
\young(12,3),
\young(13,2),
\young(21,3),
\young(23,1),
\young(31,2),
\young(32,1)} we find:
\begin{align*}
e_{(2,1)}&=\frac{2^2}{(3!)^2}
\bigl\{
(\eins+(1\,2))(\eins-(1\,3))
+(\eins+(1\,3))(\eins-(1\,2))
+(\eins+(1\,2))(\eins-(2\,3))+ \\
&\qquad+(\eins+(2\,3))(\eins-(1\,2))
+(\eins+(1\,3))(\eins-(2\,3))
+(\eins+(2\,3))(\eins-(1\,3))
\bigr\} \\
&=\frac{1}{3}
\bigl\{2\eins
-(1\,2\,3)-(1\,3\,2)
\bigr\}.
\end{align*}
Up to order $p=4$ the central idempotents are given by the sum of
minimal projectors of the standard tableaux -- they are orthogonal. 

The characters in the irreducible $(m)$ representation can be computed through
\begin{equation*}
\chi_{(m)}(s)
=\frac{f_{(m)}}{p!}\sum_{T|(m)(T)=(m)}
\sum_{\substack{p\in\Rcal(T)\\ q\in\Ccal(T) \\pq=s}}\sgn(q).
\end{equation*}
Many other useful formulas can be derived from the Frobenius character
formula.  Interchanging rows and columns in a frame $(m)$ leads us to
the \emph{dual frame} $\widetilde{(m)}$. For the characters one finds:
$\chi_{\widetilde{(m)}}(s)=\sgn(s)\chi_{(m)}(s)$. There is a nice
formula for the characters of the transpositions in \cite{bk:fulton}:

Define the rank $r$ of a frame to be the length of the diagonal. Let $a_i$
and $b_i$ be number of boxes below and to the right of the $i$'th box, reading
from lower right to upper left. Call 
$\begin{pmatrix}a_1&\dots&a_r\\b_1&\dots&b_r\end{pmatrix}$ the
characteristics of $(m)$, e.g.
\[
\young(X\hfil\hfil\hfil\hfil\hfil,%
       \hfil X\hfil\hfil,%
       \hfil\hfil X,%
       \hfil\hfil,%
       \hfil\hfil)
\quad r=3,
\text{ characteristics} =
\begin{pmatrix} 0&3&4 \\ 0&2&5 \end{pmatrix}.
\]
Then
\[
\chi_{(m)}(\tau)=
\frac{f_{(m)}}{p(p+1)}\sum_{i=1}^r(b_i(b_i+1)-a_i(a_i+1)).
\]

\bibliography{literatur}
\end{document}